\documentclass[aps,prb,preprint,showpacs]{revtex4-1}
\usepackage[dvipdfm]{graphicx,color}
\usepackage{amsmath, amssymb}
\usepackage{longtable}
\usepackage{natbib}

\begin{document}
\title{Critical Reinvestigation on Vibronic Couplings in Picene\\
from View of Vibronic Coupling Density Analysis}
\author{Tohru Sato}
\email[]{tsato@moleng.kyoto-u.ac.jp}
\affiliation{Department of Molecular Engineering, Graduate School of Engineering,
Kyoto University, Kyoto 615-8510, Japan}
\author{Naoya Iwahara}
\affiliation{Department of Molecular Engineering, Graduate School of Engineering,
Kyoto University, Kyoto 615-8510, Japan}
\author{Kazuyoshi Tanaka}
\affiliation{Department of Molecular Engineering, Graduate School of Engineering,
Kyoto University, Kyoto 615-8510, Japan}
\date{\today}

\begin{abstract}
Vibronic coupling constants  
in the monoanionic, trianionic, and excited states 
of picene are evaluated from the total energy gradients 
using the density functional theory.
Employing the calculated 
vibronic coupling constants
in the excited state 
of the neutral molecule,
electron energy loss spectrum (EELS)
is simulated to be compared with the experimental spectrum.
The calculated 
vibronic coupling constants
are analyzed 
in terms of the vibronic coupling density 
which enables us to analyze vibronic couplings 
based on the relation between the electronic and vibrational structures.
The vibronic coupling constants
reported by Kato {\it et al.} 
[{\em J. Chem. Phys.} {\bf 116}, 3420(2002) and 
{\em Phys. Rev. Lett.} {\bf 107}, 077001 (2011)] are critically discussed
based on the vibronic coupling density analysis.
\end{abstract}

\maketitle
%
%

After the discovery of the superconductivity 
in alkali-metal (K, Rb) doped picene, \cite{Mitsuhashi2010a}
experimental and theoretical studies on the electronic structure on the picene
have been piled up.
\cite{Okazaki2010a, Roth2010a, Roth2011a, Roth2011b, 
Kosugi2009a,Giovannetti2011a,Kim2011a,deAndres2011a,Subedi2011a,
Kato2011a,Casula2011a,Kubozono2011a, Kosugi2011b}
The vibronic coupling (electron-vibration coupling)
\cite{Bersuker1989a} is an important interaction 
in the electronic properties such as superconductivity.
Okazaki {\it et al}. have discussed an importance of the vibronic couplings in 
doped picene based on their photoelectron spectra (PES). \cite{Okazaki2010a}
Therefore, evaluation of the vibronic coupling constants (VCC) is
crucial to discuss electronic properties 
of doped picenes.
Some authors have published the VCCs of 
picene anions. \cite{Kato2002a,Kato2003a,Kato2011a,Subedi2011a}
However, some of the calculated VCCs are controversial.
Subedi and Boeri 
have concluded that 
the electron-phonon coupling of the modes around 1600 cm$^{-1}$ are strong,
\cite{Subedi2011a}
while those calculated by Kato {\it et al} are weak in this region.
\cite{Kato2002a, Kato2003a, Kato2011a}

Vibronic effects can be experimentally observed in spectra. \cite{Bersuker1989a}
Roth {\it et al}. measured 
electron energy loss spectrum (EELS) of pristine picene 
at 20 K. \cite{Roth2011b}
They have observed vibronic progressions 
in the EELS of the intramolecular excitation to $S_2(^1B_2)$ state. 

We have recently published calculation
of the VCCs in C$_{60}^{-}$
from the gradients of the total energies. \cite{Iwahara2010a}
The results are consistent with the experimental observation 
of the PES by Wang {\it et al.} \cite{Wang2005a}
We have proposed a concept, {\it vibronic coupling density} (VCD).
\cite{Sato2008a,Sato2009a}
Based on the VCD, we can discuss vibronic couplings from view 
of electronic and vibrational structures.
On the basis of the VCD analysis, 
we have succeeded in designing carrier-transporting molecules
with small vibronic couplings which is required in organic electronics
such as organic light-emitting diodes (OLED).
\cite{Shizu2010a,Shizu2011a}

In this work, we report the VCCs of the excited state of the free molecule 
in the neutral state,
the monoanionic, and the trianionic states of the free molecule 
based on the same method of calculation
employed in the calculation for C$_{60}^{-}$.
Using the VCCs of the excited $^1B_2$ state, 
we simulate EELS and compare the spectrum with the experimental one.
\cite{Roth2011b}
From the view of the VCD analysis, 
we critically discuss the previous VCCs
in Refs. \onlinecite{Kato2002a,Kato2003a,Kato2011a}.

%
%

%
%
We evaluated VCCs of mode $\alpha$ $V_{\alpha}$
from the gradients of the adiabatic potential energy surface $E$ 
with respect to a mass-weighted normal coordinates $Q_{\alpha}$:
\cite{Iwahara2010a,Shizu2010a,Shizu2011a,Sato2011a}
\begin{equation}
V_{\alpha}
=
\left\langle\Psi\left|
\left(\frac{\partial {\hat H}}{\partial Q_{\alpha}}\right)_{\mathbf{R}_{0}}
\right|\Psi\right\rangle
=
\left(\frac{\partial E}{\partial Q_{\alpha}}\right)_{\mathbf{R}_{0}}
,
\label{Eq:VCC}
\end{equation}
where ${\hat H}$ denotes a molecular Hamiltonian, 
$\mathbf{R}_{0}$ is the equilibrium geometry of the ground state of the neutral picene,
$\Psi$ is an electronic wavefunction of the excited or anionic state at $\mathbf{R}_{0}$,
and the phase of a normal mode $\alpha$ is chosen so 
that $V_\alpha$ becomes negative.
The vibronic Hamiltonian is written as
\begin{equation}
{\hat H}_{\rm vibro}
=
\sum_{\alpha}
\left[
{\hat T}(Q_{\alpha})
+
\frac{1}{2}\omega_{\alpha}^2Q_{\alpha}^{2}
+
V_{\alpha} Q_{\alpha}
\right]
,
\label{Eq:VibronicH}
\end{equation}
where ${\hat T}(Q_{\alpha})$ denotes kinetic energy operator of a vibration and 
$\omega_{\alpha}$ vibrational frequency.
We employed Becke's hybrid functional (B3LYP) 
\cite{Becke1993a} 
and Perdew and Wang's one with generalized gradient approximation (PW91)
\cite{Perdew1992a}
with the triple-zeta 6-311+G(d,p) basis set.
The geometries were optimized for the neutral ground state.
The optimized structures with $C_{2v}$ symmetry 
were checked with vibrational analysis to be a minimum.
The time-dependent density-functional-theory 
is applied for the excited state calculations.
We performed analytical force-calculations
for the excited and the anionic states to obtain the VCCs.\cite{Frisch2010}
The electronic and vibrational structures as well as the forces were obtained 
using a program package, Gaussian 09.\cite{Frisch2010} 
The VCCs are calculated using our codes.

\begin{figure}
\begin{center}
\begin{tabular}{cc}
\multicolumn{1}{l}{{\bf (a)} PW91}
&
\multicolumn{1}{l}{{\bf (b)} B3LYP}
\\
\includegraphics[width=0.35\hsize, angle=-90]{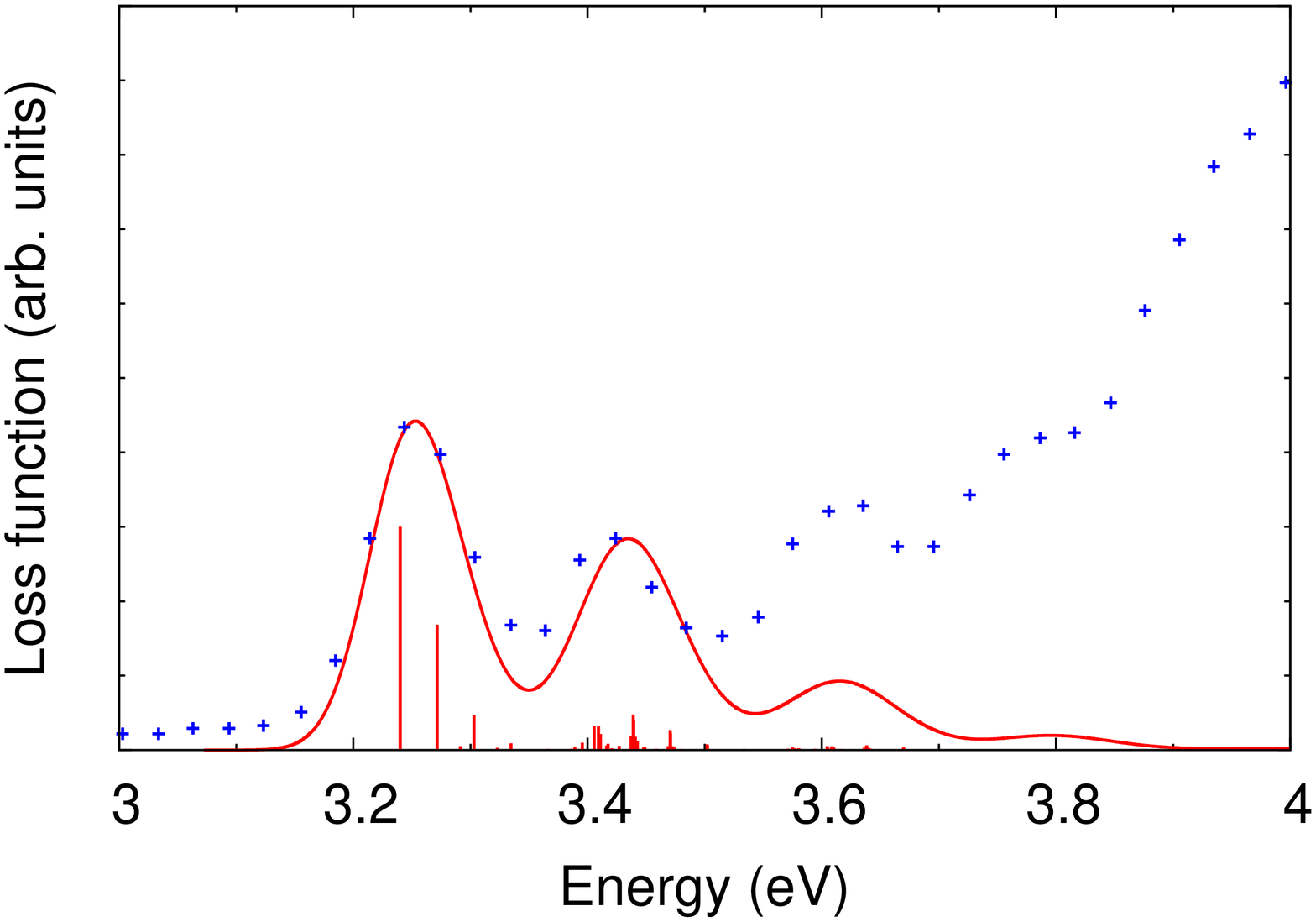}
&
\includegraphics[width=0.35\hsize, angle=-90]{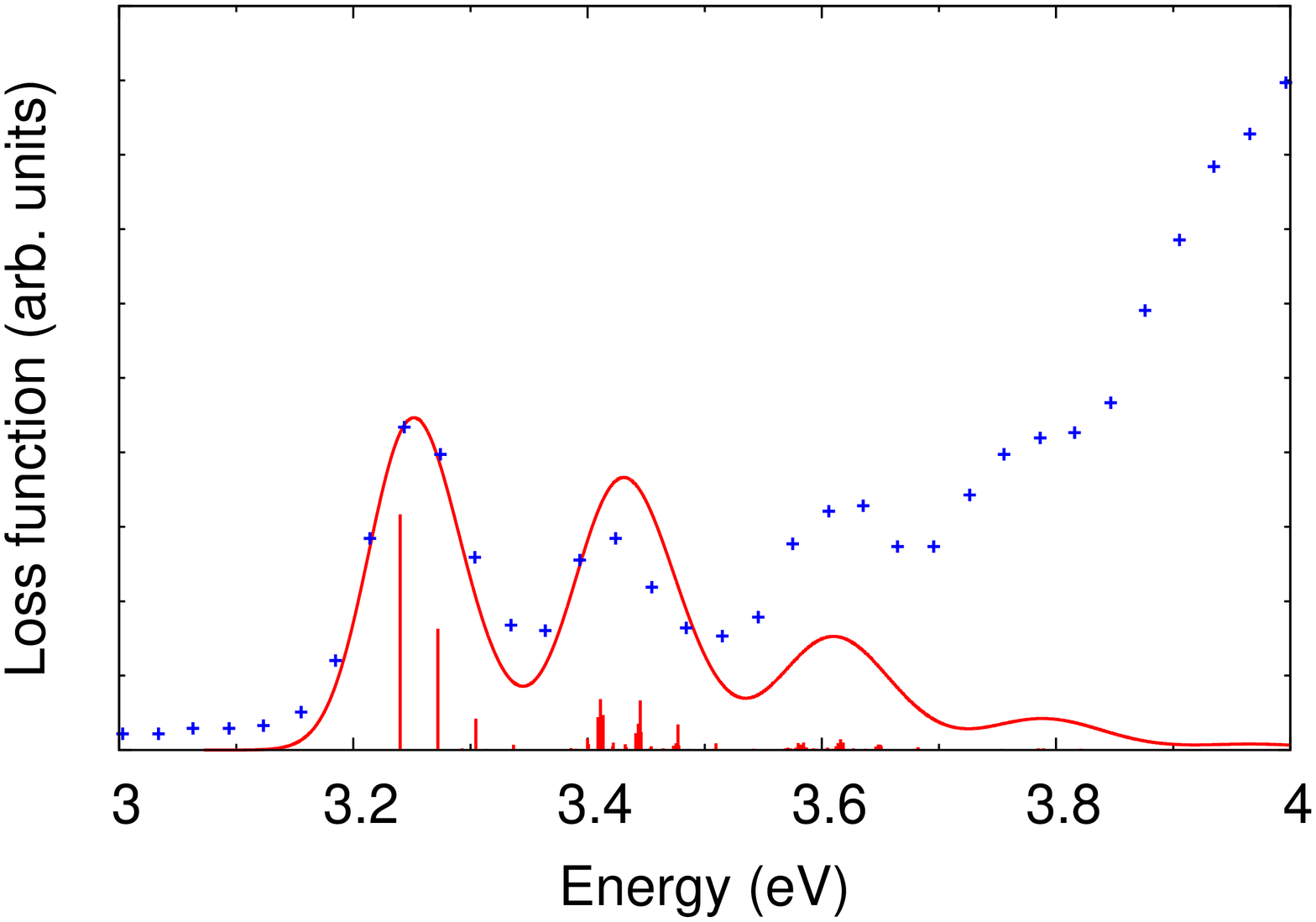}
\end{tabular}
\end{center}
\caption{(Color online) Red lines are simulated electron energy loss spectra 
(EELS) calculated by (a) the PW91 and (b) the B3LYP functional.
Blue dots indicate the experimental one by Roth {\it et al.} \cite{Roth2011b}
The 0-0 transition is set to 3.24 eV.
The other parameters employed in the simulations are $T=20$ K and 
$\sigma=270$ cm$^{-1}$.
\label{Fig:EELS}}
\end{figure}

Since the fine structure observed in the EELS 
by Roth {\it et al}. \cite{Roth2011b}
can originate from the vibronic couplings
which they did not take into account in Ref. \onlinecite{Roth2010a}, 
we simulated the EELS to find an appropriate functional 
for the VCC calculations.
EELS was simulated employing the same method 
as described in Ref. \onlinecite{Iwahara2010a}.
We considered thermal excitation at 20K
where Roth {\it et al.} observed the EELS.
The calculated excitation energies are 3.7040 and 3.2520 eV 
for the B3LYP and PW91 functionals, respectively.
The result using the PW91 functional reproduces well
the experimental excitation energy 3.25 eV. 
Using the calculated VCCs 
in the $S_2(^1B_2)$ state, we simulated the EELS (FIGs. \ref{Fig:EELS}). 
In the simulations, the 0-0 transition is set to 3.24 eV, 
and the linewidth $\sigma$ is assumed to be 270 cm$^{-1}$ 
(The FWHM is 39.4 meV).
The calculated VCCs \cite{selection} 
as well as the vibrational frequencies
are tabulated in Supporting Information.
The spectrum calculated employing the PW91 functional (FIG. \ref{Fig:EELS}(a)) 
shows a better fit than that using the B3LYP functional
(FIG. \ref{Fig:EELS}(b)).
In the calculation employing the B3LYP, 
the second strongest vibronic couplings around 1350 cm$^{-1}$ are
estimated larger than those in the result using the PW91.
\cite{screening}
Therefore, the relative intensities of EELS in the lower energy 
are reproduced employing the PW91. \cite{EELS}
Judging from the calculations of the excitation energies and
the simulated spectra,
we employ the PW91 functional hereafter
for the calculation of the ionic states.
The further comparisons of functionals are out of the scope in the present work.

%
%

\begin{figure}
\begin{center}
\begin{tabular}{cc}
\multicolumn{1}{l}{{\bf (a)} Monoanion}
&
\multicolumn{1}{l}{{\bf (b)} Trianion}
\\
\includegraphics[width=0.35\hsize, angle=-90]{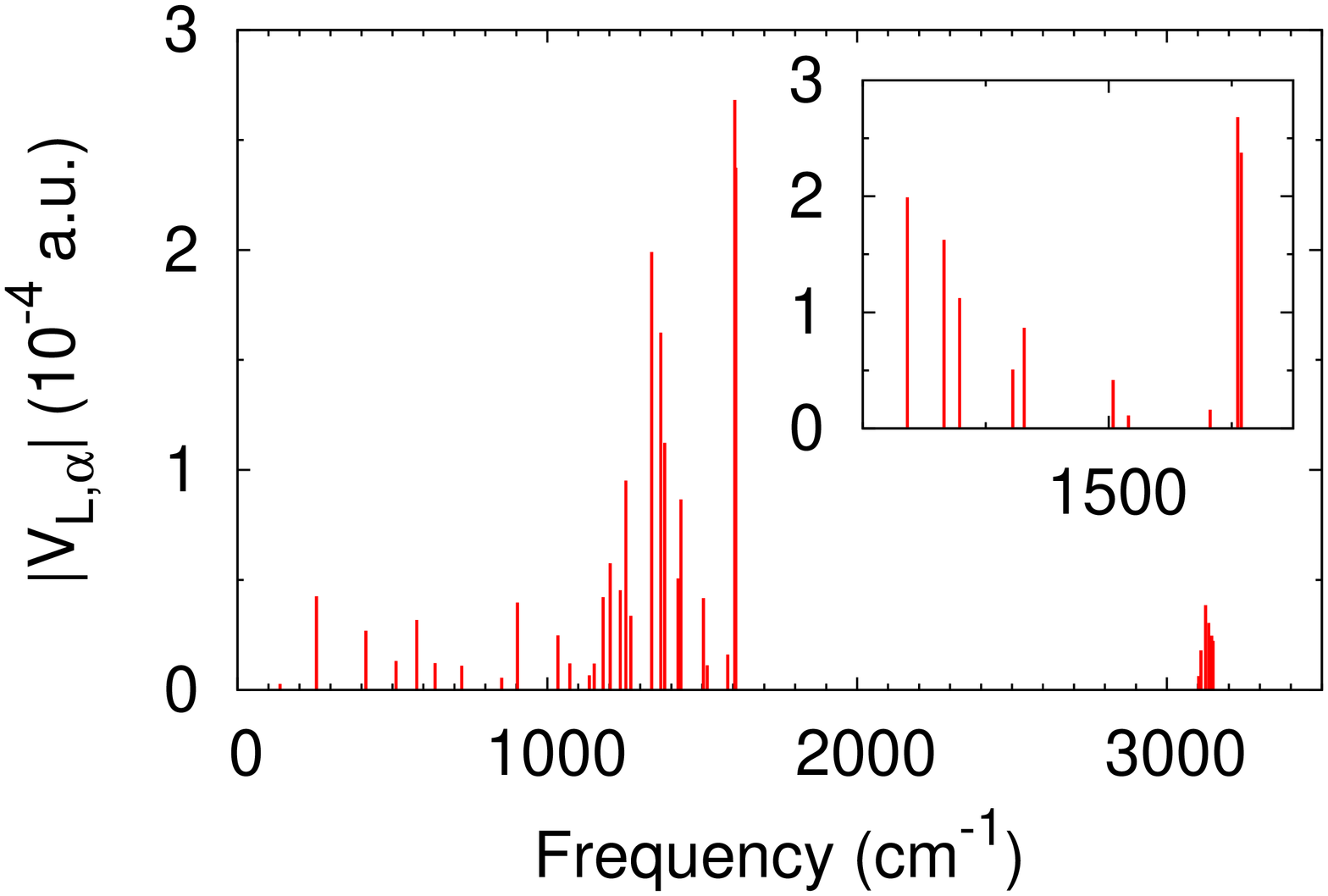}
&
\includegraphics[width=0.35\hsize, angle=-90]{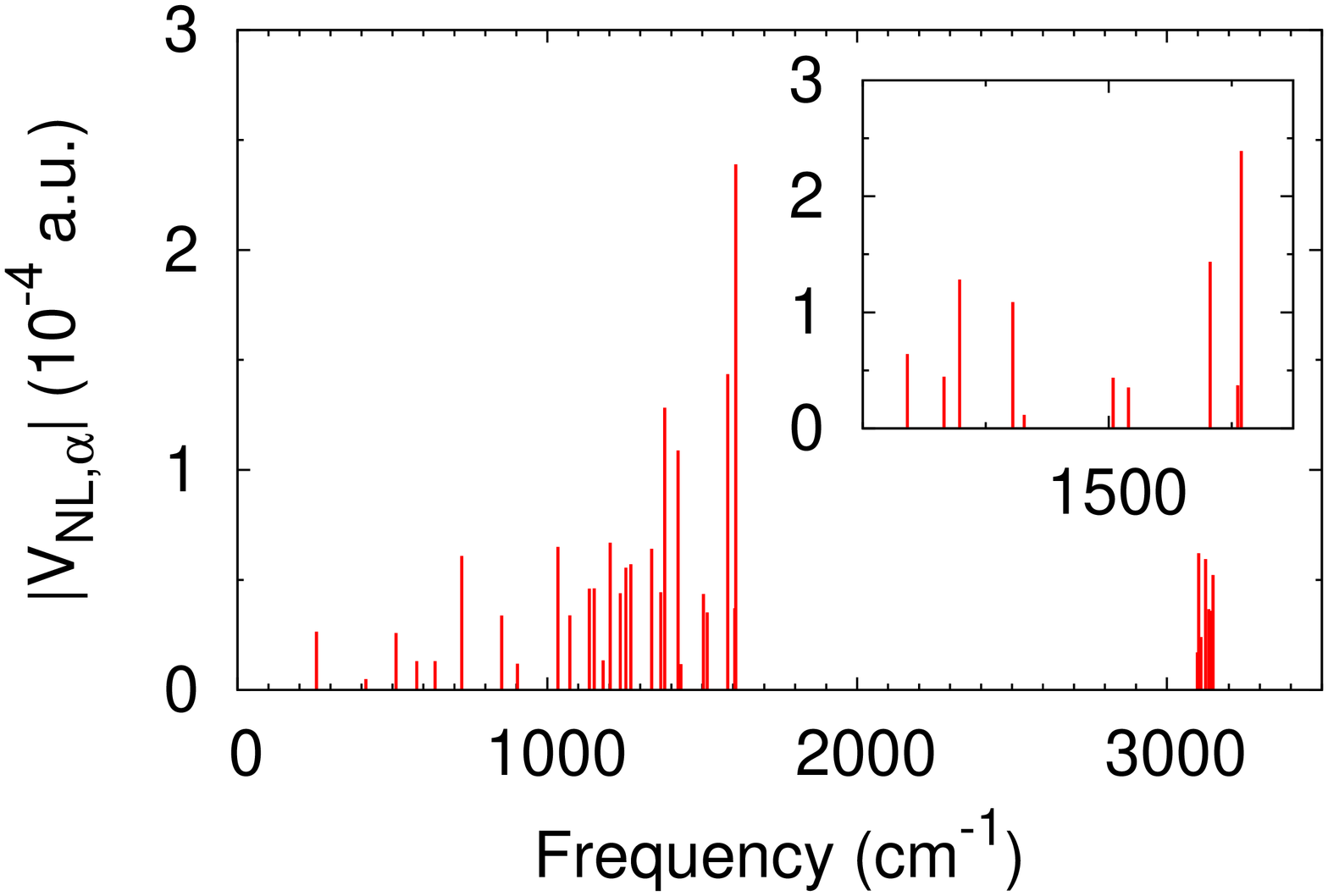}
\\
\includegraphics[width=0.35\hsize, angle=-90]{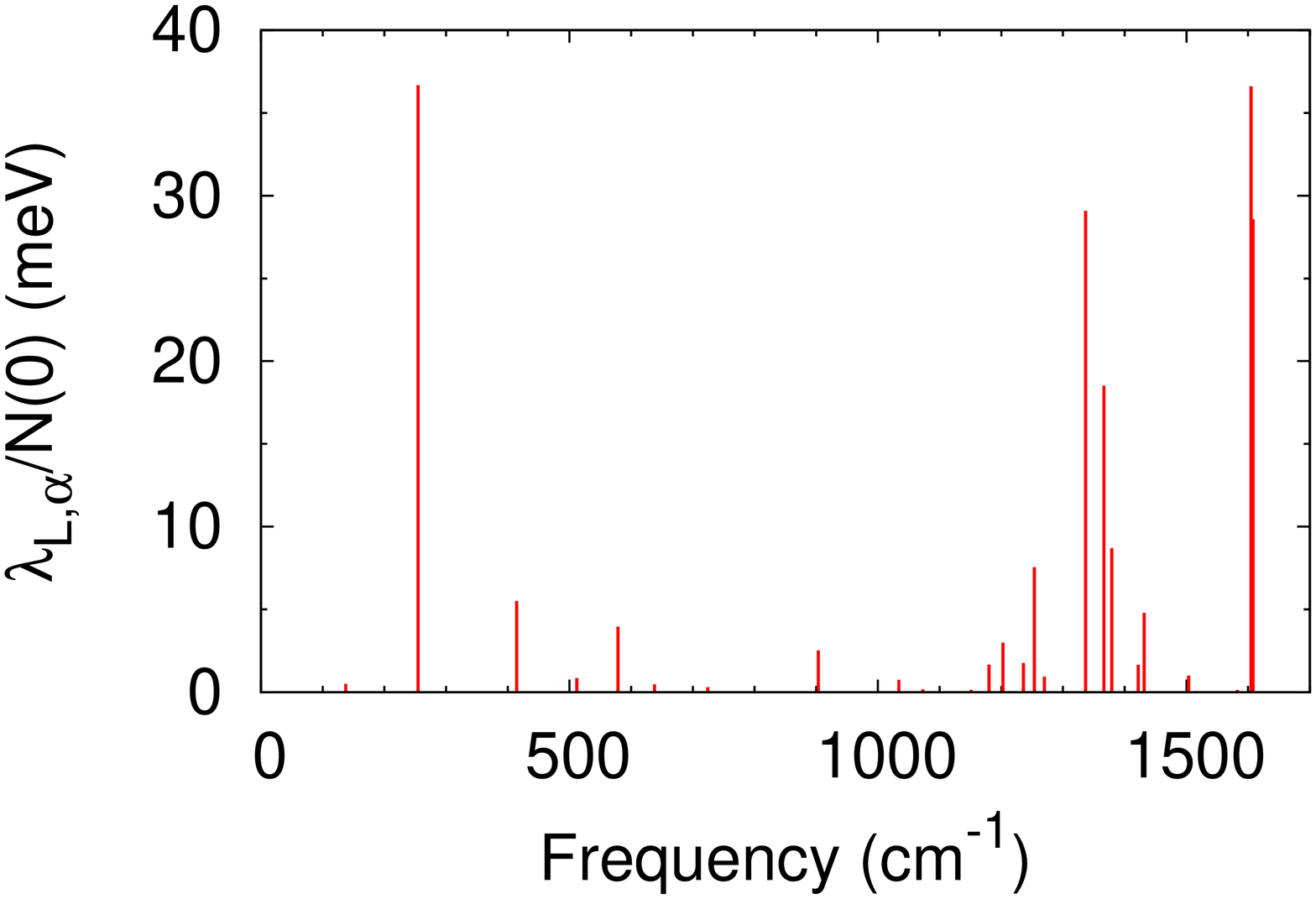}
&
\includegraphics[width=0.35\hsize, angle=-90]{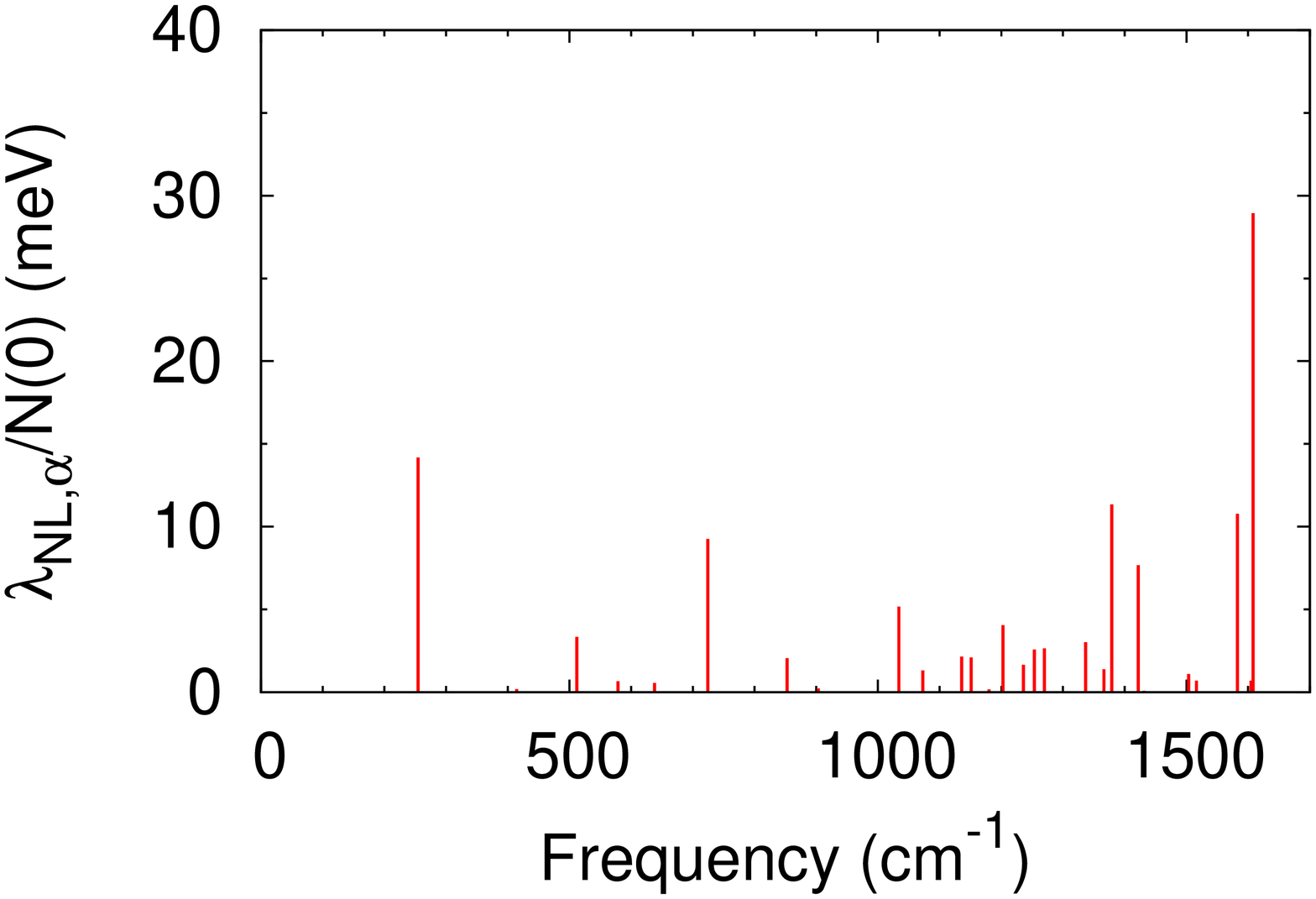}
\\
\end{tabular}
\end{center}
\caption{(Color online) Calculated (top) orbital vibronic coupling constants 
$V_{i,\alpha}$ and (bottom) electron-phonon couplings 
$\lambda_{i,\alpha}/N(0)$ in the (a) monoanion and the (b) trianion.
The subscripts L and NL denote LUMO and NLUMO, respectively.
Insets show the orbital vibronic couplings between 1300 to 1650 cm$^{-1}$.
\label{Fig:VCC-anions}}
\end{figure}

We calculated the VCCs $V_{\alpha}$
in the monoanion ($^2A_2$), dianion ($^1A_1$), and trianion ($^2B_1$) 
(see Supporting Information).
Kato {\it et al.} have calculated orbital vibronic coupling constants (OVCC)
of the lowest unoccupied molecular orbitals (LUMO) and next LUMO (NLUMO).
\cite{Kato2002a,Kato2003a,Kato2011a}
Hence we also calculated the OVCCs $V_{i,\alpha}$ from the VCCs $V_\alpha$
for comparison with the ones previously reported.
The vibronic Hamiltonian (\ref{Eq:VibronicH}) 
is mapped onto a model Hamiltonian.
The model Hamiltonian 
considered in Refs. \onlinecite{Subedi2011a} and
\onlinecite{Kato2011a} is written as follows:
\begin{eqnarray}
\hat{H}_{\rm vibro} 
=
\sum_{\alpha,i,\sigma}
\hbar \omega _{\alpha}
\left[
\hat{b}_{\alpha}^{\dagger} \hat{b}_{\alpha}
+
\frac{g_{i,\alpha}}{\sqrt{2}}
\left(
\hat{b}_\alpha^\dagger + \hat{b}_\alpha
\right)
\hat{c}_{i\sigma}^{\dagger} \hat{c}_{i\sigma}
\right]
,
\label{Eq:Hvibro}
\end{eqnarray}
where orbitals $i$ are LUMO (L) and NLUMO (NL), and 
$\alpha$ runs over all the active $a_1$ modes. \cite{selection}
The dimensionless  OVCC $g_{i,\alpha}$ is defined by
$g_{i,\alpha}=V_{i,\alpha}/\sqrt{\hbar\omega_{\alpha}^3}$.
$\hat{c}^{\dagger}_{i\sigma} (\hat{c}_{i\sigma})$ is 
the creation (annihilation) operator
of orbital $i$ and spin $\sigma$, and 
$\hat{b}^\dagger_\alpha (\hat{b}_\alpha)$ is the creation (annihilation) operator
of mode $\alpha$.
The OVCCs of LUMO $V_{{\rm L},\alpha}$ and NLUMO $V_{{\rm NL},\alpha}$ are obtained 
from the VCCs of monoanion $V_{{\rm mono},\alpha}$
and the difference between VCCs of trianion $V_{{\rm tri},\alpha}$
and dianion $V_{{\rm di},\alpha}$, respectively:
$V_{{\rm L},\alpha} = V_{{\rm mono},\alpha}$ and 
$V_{{\rm NL},\alpha} = V_{{\rm tri},\alpha} - V_{{\rm di},\alpha}$.
It should be noted that the present OVCCs effectively incorporate the contributions
from all the occupied orbitals which are important in quantitative arguments. \cite{Iwahara2010a}
The OVCCs $V_{i,\alpha}$ and intramolecular electron-phonon couplings
$\lambda_{i,\alpha}/N(0) = V_{i,\alpha}^2/\omega_\alpha^2$ 
are shown in FIGs. \ref{Fig:VCC-anions} and tabulated in Supporting Information.
$N(0)$ is the density of states at the Fermi level.

In both the monoanion and trianion, 
two sets of modes with strong VCCs are observed:
(1) the maximal coupling modes around 1600 cm$^{-1}$ 
and
(2) the second group of strong modes around 1350 cm$^{-1}$.
This is qualitatively consistent with the calculations by 
Subedi and Boeri.  \cite{Subedi2011a, SubediCalc}
However, in the calculations of Refs. 
\onlinecite{Kato2011a, Kato2002a, Kato2003a}, \cite{KatoCalc}
the modes around 1600 cm$^{-1}$ are weak
for both the monoanion and trianion.
The total electron-phonon couplings 
$\lambda_{\rm L}/N(0) = \sum_\alpha \lambda_{{\rm L}, \alpha}/N(0)$ and 
$\lambda_{\rm NL}/N(0) = \sum_\alpha \lambda_{{\rm NL}, \alpha}/N(0)$
are 196.6 meV and 119.9 meV, respectively.
The total coupling of the trianion $\lambda_{\rm NL}/N(0)$ is 
in line with that of Subedi and Boeri
(110 $\pm$ 5 meV). \cite{Subedi2011a}
On the other hand, 
in Ref. \onlinecite{Kato2011a},
$\lambda_{\rm L}/N(0) = 178$ meV and 
$\lambda_{\rm NL}/N(0) = 206$ meV.
They underestimated and overestimated the total couplings for the monoanion and trianion,
respectively.

We will discuss the disagreement
from view of the electronic and vibrational structures.
The calculated VCCs can be rationalized based on the vibronic coupling density.
\cite{Sato2008a, Sato2009a}
A VCD $\eta_{\alpha}$ is defined as
\begin{equation}
\eta_{\alpha}(\mathbf{r})
=
\Delta\rho(\mathbf{r}) \times v_{\alpha}(\mathbf{r})
,
\label{Eq:VCD}
\end{equation}
where $\Delta\rho(\mathbf{r}) =\rho(\mathbf{r})-\rho_0(\mathbf{r})$ is
the electron density difference
between the electron density of an ionic state $\rho$ and
that of a neutral state $\rho_0$.
The potential derivative $v_{\alpha}(\mathbf{r})$ is 
the derivative with respect to a mass-weighted normal coordinate $Q_{\alpha}$
of the potential $u(\mathbf{r})$ acting on one electron at a position $\mathbf{r}$
from all the nuclei.
The vibronic coupling constant is equal to the integral
of $\eta_{\alpha}(\mathbf{r})$ over space $\mathbf{r}$:
\begin{equation}
V_{\alpha}
=
\int d^3 \mathbf{r} \, \eta_{\alpha}(\mathbf{r})
.
\label{Eq:VCC-VCD}
\end{equation}
The VCD gives a local picture of the vibronic coupling,
and hence enables us to discuss the strength of the coupling qualitatively.

\begin{figure}
\begin{center}
\begin{tabular}{cc}
\multicolumn{1}{l}{\bf (a) $a_1(27)$ (1605 cm$^{-1}$)} & 
\multicolumn{1}{l}{\bf (b) $a_1(21)$ (1379 cm$^{-1}$)}
\\
\includegraphics[width=0.45\hsize]{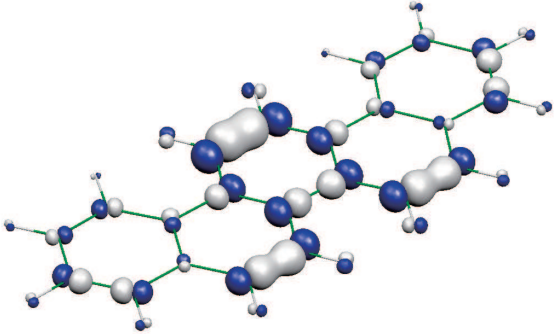}
&
\includegraphics[width=0.45\hsize]{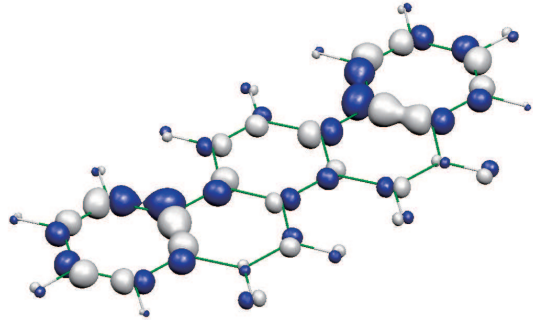}
\hspace{0.45\hsize}
\end{tabular}
\end{center}
\caption{(Color online) Potential derivatives $v_{\alpha}$
for (a) $a_1(27)$ (1605 cm$^{-1}$) and (b) $a_1(21)$ (1379 cm$^{-1}$) modes.
White and blue areas indicate positive and negative, respectively.
The threshold is 1.0 $\times$ 10$^{-2}$ a.u.
\label{Fig:Ven}}
\end{figure}

We will concentrate on the $a_{1}(27)$ mode of the maximal-coupling mode 
around 1600 cm$^{-1}$ and
the $a_{1}(21)$ mode from the second group around 1350 cm$^{-1}$
because the $a_1(27)$ and $a_1(21)$ modes are close to 
the mode $5$ in Ref. \onlinecite{Subedi2011a} and the 21st mode ($\nu_{21}$) 
in Ref. \onlinecite{Kato2011a}, respectively. 
FIGs. \ref{Fig:Ven} show the potential derivatives $v_{\alpha}(\mathbf{r})$ 
for the $a_{1}(27)$ and the $a_{1}(21)$ modes.
The distribution of the potential derivative with respect to the $a_{1}(27)$ mode is located 
on the armchair-edges of the central three hexagons.
On the other hand, that with respect to the $a_{1}(21)$ mode is on the terminal hexagons.

\begin{figure}
\begin{center}
\begin{tabular}{cc}
\multicolumn{1}{l}{\bf (a) LUMO} & \multicolumn{1}{l}{\bf (b) $\Delta\rho$}
\\
\includegraphics[width=0.45\hsize]{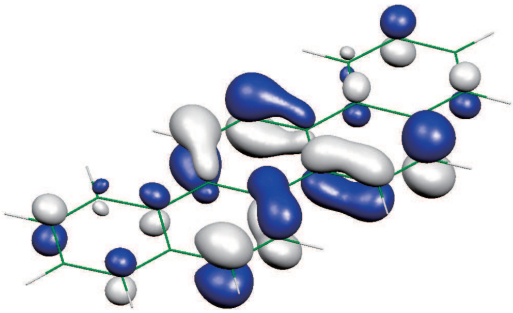}
&
\includegraphics[width=0.45\hsize]{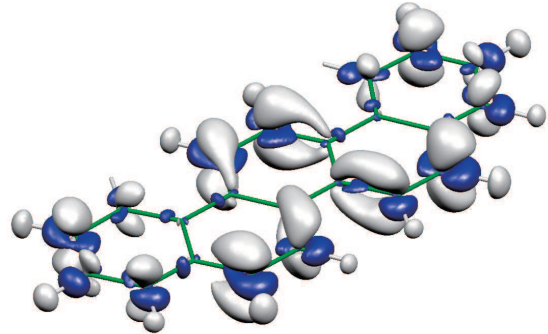}
\end{tabular}
\end{center}
\caption{(Color online) (a) LUMO and 
(b)electron density difference $\Delta \rho$ for the monoanion. 
White and blue areas indicate positive and negative, respectively.
The threshold is 5.0 $\times$ 10$^{-2}$ a.u. for LUMO and 
1.8 $\times$ 10$^{-3}$ a.u. for $\Delta \rho$. 
\label{Fig:Electronic}}
\end{figure}

FIG. \ref{Fig:Electronic}(b) shows the electron density difference 
$\Delta \rho$ of the monoanion.
Since an additional electron occupies the LUMO (FIG. \ref{Fig:Electronic}(a)), 
the positive $\pi$ density (white) appears in $\Delta \rho$.
It should be noted that there occurs decrease of the $\sigma$ density 
(blue) in the molecular plane. 
Such a polarized density originates from the Coulomb interactions 
between the electron occupying the LUMO and all electrons in doubly occupied 
orbitals below the highest occupied molecular orbital (HOMO). \cite{Sato2008a}

\begin{figure}
\begin{center}
\begin{tabular}{ll}
\multicolumn{1}{l}{\bf (a) $a_1(27)$ (1605 cm$^{-1}$)} & 
\multicolumn{1}{l}{\bf (b) $a_1(21)$ (1379 cm$^{-1}$)}
\\
\includegraphics[width=0.45\hsize]{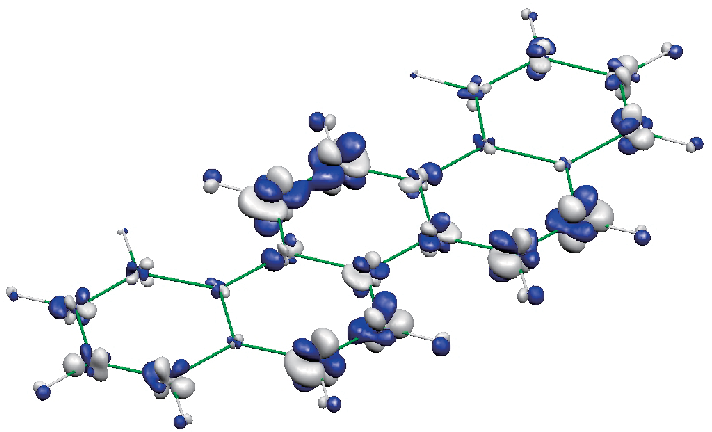}
&
\includegraphics[width=0.45\hsize]{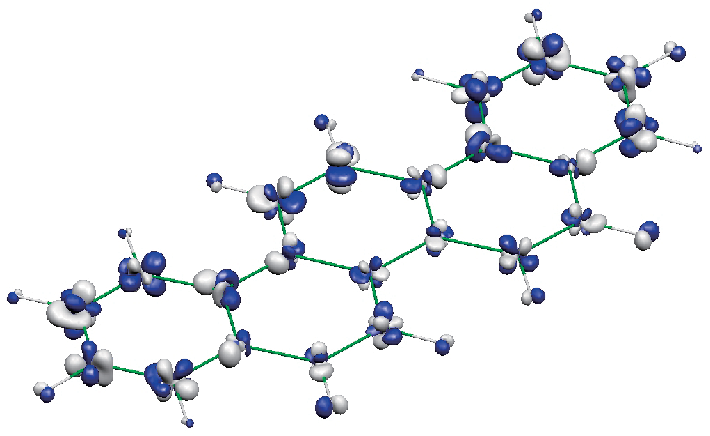}
\end{tabular}
\end{center}
\caption{(Color online) Vibronic coupling densities 
for (a) $a_1(27)$ (1605 cm$^{-1}$) and (b) $a_1(21)$ (1379 cm$^{-1}$) modes of the monoanion.
White and blue areas indicate positive and negative, respectively.
The threshold is 2.0 $\times$ 10$^{-5}$ a.u.
\label{Fig:VCD-mono}}
\end{figure}

FIGs. \ref{Fig:VCD-mono} show the vibronic coupling densities 
$\eta_\alpha(\mathbf{r})$
of the monoanion with respect to the $a_{1}(27)$ and $a_{1}(21)$ modes.
As for the $a_{1}(27)$, the electron density difference 
$\Delta\rho(\mathbf{r})$ shows considerable overlap 
with the potential derivative $v_{\alpha}(\mathbf{r})$.
On the other hand, $\Delta \rho(\mathbf{r})$ does not significantly overlap 
$v_\alpha(\mathbf{r})$ for the $a_1(21)$ mode. 
Therefore, the VCD $\eta_\alpha$ (\ref{Eq:VCD}) for the $a_1(27)$ is 
larger than that for the $a_1(21)$.
Particularly, $\eta_\alpha$ for the $a_1(27)$ has the large distribution
on the central three armchair-edges. 
Accordingly, the VCC of the $a_{1}(27)$ mode is larger than that of the $a_{1}(21)$ mode.

The VCD $\eta_\alpha$ for the $a_1(27)$ mode on the bonds appears 
due to the polarization of $\Delta \rho$, thus
neglecting such a polarization can give rise to quantitative,
or sometimes qualitative, errors 
in VCC calculations based on the orbital levels. \cite{Iwahara2010a}
In addition, an electron density difference is usually different 
from the orbital density of HOMO or LUMO.
Many-body effect sometimes plays a crucial role 
on the electron density difference, and therefore vibronic couplings. 
\cite{Sato2011a}

\begin{table}
\caption{Calculated frequencies (cm$^{-1}$) and orbital gradients 
$\partial \epsilon/\partial Q$ and total-energy gradients $\partial E/\partial Q$
as vibronic coupling constants (10$^{-4}$ a.u.) for the selected modes of the monoanion.
The subscripts L and S denote the neutral LUMO and the SOMO of the anion, respectively.
\label{table:gradients}}
\begin{center}
\begin{tabular}{ccccccccc}
\hline\hline
 & Freq.
 & $\partial \epsilon_{\rm L}/\partial Q$ 
 & $\partial \epsilon_{\rm S}/\partial Q$ 
 & \multicolumn{2}{c}{$\partial E/\partial Q$}
\\
 & & & & Num. & Anal. \\
\hline
$a_1(2)$  & 254.5 & 0.372 & 0.475 & 0.426 & 0.426 \\
$a_1(19)$ &1336.4 & 2.019 & 1.889 & 1.988 & 1.990 \\
$a_1(20)$ &1366.1 & 1.872 & 1.365 & 1.639 & 1.624 \\
$a_1(21)$ &1378.8 & 0.820 & 1.335 & 1.130 & 1.123 \\
$a_1(27)$ &1604.9 & 3.070 & 2.281 & 2.706 & 2.682 \\
$a_1(28)$ &1607.9 & 2.090 & 2.580 & 2.377 & 2.374 \\
\hline\hline
\end{tabular}
\end{center}
\end{table}

Kato {\it et al}.  have calculated VCCs as the gradients of the orbital 
levels (orbital gradients) with respect to normal coordinates. \cite{Kato2002a}
We also obtained the orbital gradients of selected modes
from the neutral LUMO $\partial \epsilon_{\rm L}/\partial Q_{\alpha}$
and from the singly occupied molecular orbital (SOMO) 
of the monoanion $\partial \epsilon_{\rm S}/\partial Q_{\alpha}$
to compare the present VCCs calculated analytically 
from the total energy gradient 
as well as numerical gradients.
The orbital gradients and the numerical energy gradients were obtained
by fitting linear and quadratic polynomials, respectively,  
in the range from $-$0.2 to 0.2 a.m.u.$^{1/2}$ $a_0$ 
with a step size 0.05 a.m.u.$^{1/2}$ $a_0$ where $a_0$ is the Bohr radius.
These range and step size could be different from those of Kato {\it et al}. 
We summarize the results of the calculations in TABLE \ref{table:gradients}.
All the results indicate that 
the vibrational modes around 1600 cm$^{-1}$ have the maximal coupling.

Though the gradients of the LUMO level can yield results which 
are qualitatively consistent with the gradients of the total energy,
the results in Refs. \onlinecite{Kato2002a, Kato2011a} are not the case.
In their calculations, the vibrational mode around 1380 cm$^{-1}$ has
the maximal coupling, 2.3$\times 10^{-4}$ a.u. for the monoanion. \cite{Kato2002a}
However, since $\Delta\rho$ (see FIG.\ref{Fig:Electronic}(b)) is mainly 
located on the armchair-edges of the central three hexagons,
it does not overlap with the potential derivatives $v_\alpha$ of the mode 
(FIG.\ref{Fig:Ven}(b)), the VCC of the mode cannot be the maximal.
Similar discussion holds for the mode around 1520 cm$^{-1}$ 
which has the maximal coupling, 2.6$\times 10^{-4}$ a.u.
for the trianion. \cite{Kato2011a}

%
%
In summary,
we calculated the VCCs 
of picene 
for the excited state $^1B_2$ of the neutral molecule, 
the monoanionic, and the trianionic states of the molecule
from the gradients of the total energies 
with respect to the normal modes.
In the previous studies, 
the VCCs calculated form the gradients of the orbital energies
of the frontier level, LUMO.
In other words, 
they have regarded the OVCCs of the LUMO as the VCCs.
The present approach can provide quantitatively reliable VCCs,
since all the occupied orbitals can contribute to the VCCs
due to the selection rule.
This has been discussed in Ref. \onlinecite{Iwahara2010a}
in detail.
We simulated the EELS and compared the spectra
with experimental spectrum by Roth {\it et al}.
No comparison with experiments
has been reported on the vibronic couplings in picene.
Needless to say,
such a comparison 
between the theoretical and the experimental results is
necessary to obtain reliable VCCs.
From these simulations, we determined an appropriate functional for
the calculation of the vibronic couplings in picene.
Using the functional which reproduces the EELS,
we evaluated the vibronic coupling constants of the picene anions.
The calculated vibronic couplings of the anions can be employed
for assignments of spectra, for example, photoelectron spectra
of the anions.
We discussed the vibronic couplings of the picene anions
in terms of the vibronic coupling density.
Based on the analysis, the present vibronic couplings are reasonable
compared with the values reported previously.

%
%
Numerical calculations were performed partly
in the Supercomputer Laboratory of Kyoto University
and Research Center for Computational Science, Okazaki, Japan.
This work was supported in part
by the Japan Society for the Promotion of Science (JSPS)
through its Funding Program
for the Global COE Program
``International Center for Integrated Research
and Advanced Education in Materials Science'' (No. B-09)
of the Ministry of Education, Culture, Sports, Science and Technology (MEXT) of Japan.


\begin{thebibliography}{33}
\bibitem [{scr()}]{screening}%
  \BibitemOpen
  \href@noop {} {}\bibinfo {note} {In terms of the electron-phonon
  couplings, the total coupling in the excited state obtained by the B3LYP
  (355.8 meV) is larger than that by the PW91 (291.4 meV) by 64
  meV.}\BibitemShut {Stop}%
\end{thebibliography}

%
\end{document}